\documentclass[12pt,preprint]{aastex}

\usepackage{graphicx}
\usepackage{color}
\usepackage{amsmath,amssymb}
\usepackage{subfig}

\newcommand{\mlcs}{\textsc{mlcs2k2}}
\newcommand{\salt}{\textsc{salt2}}
\newcommand{\dbll}{$\lambda\lambda$}
\newcommand{\OII}{[O~{\footnotesize II}]}
\newcommand{\OIII}{[O~{\footnotesize III}]}
\newcommand{\NII}{[N~{\footnotesize II}]}
\newcommand{\SII}{[S~{\footnotesize II}]}
\newcommand{\tolmstead}{(Olmstead et al., in prep)}

\newcommand{\ssfrp}{sSFR$_{\textrm{phot}}$}
\newcommand{\sfrp}{SFR$_{\textrm{phot}}$}
\newcommand{\ssfrs}{sSFR$_{\textrm{spec}}$}
\newcommand{\sfrs}{SFR$_{\textrm{spec}}$}
\newcommand{\sdss}{SDSS}
\newcommand{\sdsssns}{SDSS-SNS}

\begin{document}

\title{Spectroscopic Properties of Star-Forming Host Galaxies and Type~Ia Supernova Hubble Residuals in a Nearly Unbiased Sample}

\shorttitle{SDSS-II SN Survey: SNe\,Ia Host Metallicities}
\shortauthors{D'Andrea et al.}

\author{Chris B. D'Andrea\altaffilmark{1,2},
Ravi R. Gupta\altaffilmark{1},
Masao Sako\altaffilmark{1},
Matt Morris\altaffilmark{1,3},
Robert C. Nichol\altaffilmark{2},
Peter J. Brown\altaffilmark{4},
Heather Campbell\altaffilmark{2},
Matthew D. Olmstead\altaffilmark{4},
Joshua A. Frieman\altaffilmark{5,6,7},
Peter Garnavich\altaffilmark{8},
Saurabh W. Jha\altaffilmark{9},
Richard Kessler\altaffilmark{5,6},
Hubert Lampeitl\altaffilmark{2},
John Marriner\altaffilmark{7},
Donald P. Schneider\altaffilmark{10},
Mathew Smith\altaffilmark{11}
}

\email{chris.dandrea@port.ac.uk}
\altaffiltext{1}{Department of Physics and Astronomy, University of Pennsylvania, 209 South 33rd Street, Philadelphia, PA 19104, USA}
\altaffiltext{2}{Institute of Cosmology and Gravitation, University of Portsmouth, Dennis Sciama Building, Burnaby Road, Portsmouth, PO1 3FX, UK}
\altaffiltext{3}{Department of Physics and Astronomy, Johns Hopkins University, 3400 North Charles Street, Baltimore, MD 21218, USA}
\altaffiltext{4}{Department of Physics and Astronomy, University of Utah, Salt Lake City, UT 84112, USA}
\altaffiltext{5}{Kavli Institute for Cosmological Physics, The University of Chicago, 5640 South Ellise Avenue, Chicago, IL 60637, USA}
\altaffiltext{6}{Department of Astronomy and Astrophysics, The University of Chicago, 5640 South Ellise Avenue, Chicago, IL 60637, USA}
\altaffiltext{7}{Center for Astrophysics, Fermi National Accelerator Laboratory, P.O. Box 500, Batavia, IL 60510, USA}
\altaffiltext{8}{Department of Physics and Astronomy, Rutgers University, 136 Frelinghuysen Road, Piscataway, NJ 08854, USA}
\altaffiltext{9}{Department of Physics, University of Notre Dame, 225 Nieuwland Science Hall, Notre Dame, IN 46556, USA}
\altaffiltext{10}{Department of Astronomy and Astrophysics, The Pennsylvania State University, 525 Davey Laboratory, University Park, PA 16802, USA}
\altaffiltext{11}{Astrophysics, Cosmology and Gravity Centre, Department of Mathematics and Applied Mathematics, University of Cape Town, Rondebosch 7701, Cape Town, South Africa}

\begin{abstract}
We examine the correlation between supernova host galaxy properties and their residuals on the Hubble diagram.  We use supernovae discovered during the Sloan Digital Sky Survey II - Supernova Survey, and focus on objects at a redshift of $z < 0.15$, where the selection effects of the survey are known to yield a complete Type Ia supernova sample.  To minimize the bias in our analysis with respect to measured host-galaxy properties, spectra were obtained for nearly all hosts, spanning a range in magnitude of $-23 < M_r < -17$.  In contrast to previous works that use photometric estimates of host mass as a proxy for global metallicity, we analyze host-galaxy spectra to obtain gas-phase metallicities and star-formation rates from host galaxies with active star formation.  From a final sample of $\sim40$ emission-line galaxies, we find that light-curve corrected Type Ia supernovae are $\sim 0.1$ magnitudes brighter in high-metallicity hosts than in low-metallicity hosts.  We also find a significant ($>3\sigma$) correlation between the Hubble residuals of Type Ia supernovae and the specific star-formation rate of the host galaxy.  We comment on the importance of supernova/host-galaxy correlations as a source of systematic bias in future deep supernova surveys. 
\end{abstract}


\section{Introduction}
\label{sec:intro}

The utility of Type Ia supernovae (SNe Ia) to the study of cosmology rests upon the ability to calibrate the difference in absolute magnitude between events, allowing for an accurate determination of the distance-redshift relationship.  Variations in the intrinsic luminosity of each event can be measured by the width and color of the observed light curve \citep{Phillips,Hamuy96,RPK96}.  After applying these relations using light-curve fitting codes such as \salt\ \citep{SALT2} or \mlcs\ \citep{MLCS}, the resulting derived distance moduli display an intrinsic scatter about the best-fit cosmology equivalent to $\approx 7\%$ in distance.  The difference between the measured distance modulus of a SN and the best-fit cosmology is known as the Hubble Residual (HR).

The absolute magnitude of SNe Ia have additionally been shown to depend on their environment; SNe Ia in late-type galaxies are intrinsically more luminous than those in early-type hosts \citep{Hamuy96,Gallagher05}.  One possible explanation for this difference is the progenitor metallicity.  \citet{Timmes} showed, both analytically and through modelling, that a metal-rich progenitor produces less $^{56}$Ni, and as such is less luminous, than a metal-poor SN Ia progenitor.  If this difference in luminosity as a function of metallicity is accompanied by the same changes in color and light-curve stretch that normally occur in SNe Ia, then this environmental effect would not produce biased distance measurements.  However, \citet{KRW} have recently shown that light-curve corrections tend to overcompensate for the metallicity effect.  Specifically, they show that for a given quantity of synthesized $^{56}$Ni, multi-dimensional models predict that a higher metallicity progenitor will produce a narrower light curve.  This effect means that after standard light-curve corrections are applied, a high-metallicity progenitor would appear to be overluminous.

Several groups have recently shown that the host galaxy environment does in fact correlate with the HR, in the sense that more massive galaxies host overluminous (for their light-curve shape) SNe Ia \citep{Gallagher08,Kelly10,Sullivan10,Lampeitl10}.  Under the assumption that these galaxies all follow the mass-metallicity relationship derived in \citet[hereafter T04]{Tremonti}, where high-mass galaxies have high metallicity, \citet{Sullivan10} showed that their results are consistent with the metallicity/luminosity relation in \citet{KRW}.  

Knowing the true cause of the HR correlation with host galaxy properties is key for future deep SN surveys such as the Dark Energy Survey\footnote{http://www.darkenergysurvey.org/} (DES) and the Large Synoptic Survey Telescope \citep[LSST;][]{LSST}.  While it is clear that correcting for this relationship would lead to decreased scatter in HRs, care must be taken to consider the change in galaxy properties with redshift.  If progenitor metallicity truly is the cause of the observed correlation, and host mass is used as a proxy for the metallicity, then evolution in the mass-metallicity relationship as a function of redshift \citep{Maiolino} will bias the Hubble Diagram and our measurement of cosmological parameters.  The correlation might instead be due to the relative prevelance of progenitor channels and thus depend on star-formation rate (SFR), which will have a different redshift dependence.  \citet{Gallagher08} showed a correlation exists between stellar metallicities from passive galaxies and SN Ia HRs at low redshifts ($z < 0.05$), but no correlation between measured emission-line metallicities from star-forming galaxies (SFG) has yet been observed \citep[see][]{Gallagher05}.  Furthermore, other factors may contribute part of the intrinsic scatter, such as explosion geometry \citep{Maeda}, without being directly tied to host-galaxy properties.

In what follows we determine the relationship between HRs and both host galaxy metallicity and SFRs in a nearly unbiased sample of spectroscopically-observed host galaxies of SNe Ia discovered as part of the {\sc SDSS-II} Supernova Survey.  Section~\ref{sec:data} describes our observations and the construction of our sample.  We discuss our method for measuring the gas-phase metallicity and specific SFR (SFR per unit stellar mass; sSFR) of star-forming host galaxies of SNe Ia in Section~\ref{sec:late}.  Results are presented in Section~\ref{sec:results}, and the implications of our findings for SN Ia cosmology are detailed in Section~\ref{sec:discuss}.  

\section{Observations}
\label{sec:data}

We apply our analysis to host galaxies of the SNe Ia discovered as part of the Sloan Digital Sky Survey-II Supernova Survey \citep[hereafter \sdsssns;][]{Frieman08}.  The \sdsssns\ repeatedly surveyed the 300 square degree Southern Equatorial Stripe \citep[designated stripe 82; see][]{Stoughton} during the Fall seasons (September 1 - November 30) of 2005-2007 using the dedicated 2.5 meter SDSS telescope at Apache Point Observatory, New Mexico \citep{SDSS-Telescope}.  Each photometric observation consists of nearly simultaneous 55 second exposures in each of the five \emph{ugriz} filters \citep{SDSS-Filters} using the wide-field \sdss\ CCD camera \citep{SDSS-Camera}.  High quality light curves were obtained \citep{JH08} on a photometric system calibrated to an uncertainty of 1\% \citep{Ivezic}.  For a technical summary of the \sdss, see \citet{York}.

The \sdsssns\ spectroscopically confirmed 504 SNe Ia over the duration of its three year survey.  Details of the spectroscopic observing campaign and the criteria by which targets were selected can be found in \citet{Sako08}.  An additional 210 transients with identifiable hosts have been designated photometrically probable SNe Ia.  These objects did not have a spectrum taken during the survey, usually because they either (a) were not well separated from their host, or (b) were below the magnitude limit for our followup spectroscopy.  However, these objects are highly likely to be SNe Ia as opposed to any of the core-collapse types based on their multicolor light curves and the redshifts obtained from spectra of their host galaxies.  The updated photometric classifier code, containing more core-collapse templates and described in \citet{Sako11}, was designed for the SDSS-SNS and has been shown to be accurate at determining SNe Ia with low contamination \citep{Kessler10}.  Where necessary we will refer to the spectroscopically-confirmed sample as Spec Ia SNe and the photometrically-probable sample with identified hosts as Phot Ia SNe.

We derive distance moduli from our sample of SNe Ia with the SNANA code \citep{Kessler09b}, using both the \mlcs\ and \salt\ light-curve fitters.  Light-curve quality cuts applied are the same as those in Section 4 of the SDSS-SNS first-year cosmology paper \citep{Kessler09a}, with the one exception being a more stringent requirement of at least one measurement at two days or more before peak brightness in the rest-frame \emph{B}-band according to \mlcs\ (T$_{\textrm{rest}} < -2$).  These cuts remove SNe Ia that have low signal-to-noise measurements, insufficient temporal coverage, and peculiar light-curve shapes.  Distance moduli from \salt\ light-curve fits are computed using the code `SALT2mu' \citep{Marriner}.  The values used for the $\alpha$ and $\beta$ parameters, which are the corrections for stretch and color, are determined in \citet{Marriner} as those best-fit by the \sdsssns\ data, independent from cosmology.  For \mlcs, the reddening law $R_{V}=2.03$ is used, as is the default SNANA prior on $A_{V}$ of $\exp{(-A_V/0.3)}$.  Hubble Residuals are determined by subtracting the distance modulus of the assumed cosmology at the redshift of the host galaxy from that of the SN (HR $\equiv \mu_{\textrm{SN}}-\mu_{\textrm{z}}$).  We use the best-fit $\Lambda$CDM cosmology to the SDSS-only SN sample from \citet{Kessler09a}, $\Omega_M=0.274$ and $\Omega_{\Lambda}=0.735$. We assume a Hubble Constant of $H_0=70$ km sec$^{-1}$ Mpc$^{-1}$ throughout this paper, though we note that the choice of Hubble Constant is irrelevant to our results, as $H_0$ is degenerate with the fiducial SN Ia absolute luminosity.

A point of emphasis in our analysis is to ensure that selection biases in our data set are minimized to the greatest extent possible.  To this end we have limited our study to SNe Ia at redshifts $z < 0.15$, where it has been demonstrated that the selection efficiencies of the survey are such that the Spec Ia plus Phot Ia SNe sample is complete \citep{Dilday08,Dilday10}.  Thus our results will not be biased by containing an over-abundance of intrinsically over-luminous SNe Ia.  The galaxy spectra analyzed in this paper are hosts of the 140 Spec Ia and 7 Phot Ia SNe at $z < 0.15$, of which 77 Spec Ia and 3 Phot Ia SNe pass our light-curve quality cuts. 

A concerted effort is being undertaken to obtain galaxy-only spectra for all Spec Ia and Phot Ia SNe discovered in the SDSS-SNS.  For the low-redshift sample analyzed in this paper, this campaign is essentially complete; we have host-galaxy spectra for 93\% percent of the low-$z$ sample.  Over half of these SNe Ia have host-galaxy spectra from the SDSS Legacy Survey \citep[hereafter referred to as `SDSS spectra';][]{DR7}, and were obtained from the SDSS Data Archive Server (DAS).  In collaboration with the SDSS-III Baryon Oscillation Spectroscopic Survey \citep[BOSS;][]{Eisenstein} most of the remaining unobserved SNe Ia hosts with $m_r < 20.5$ were targeted, including over a third of the low-$z$ sample.  The faintest host galaxies were observed with the 8m Gemini-South Telescope and are vital for the completeness of our sample, as the least luminous galaxies tend to have the lowest metallicities.  The remainder of the host-galaxy spectra used in this analysis were obtained with the 3.5m Astrophysical Research Consortium Telescope at Apache Point Observatory (APO) and the 3.6m New Technologies Telescope (NTT), which observed some hosts lacking SDSS spectra before the BOSS and GEMINI observations were begun.  

The fact that we begin with a nearly complete sample of host-galaxy spectra is crucial, as this means we are not biased toward more luminous hosts.  The parameters that we are attempting to directly measure (metallicity, SFR) are correlated with the absolute magnitude of a galaxy, as demonstrated in T04: metal-poor galaxies tend to be faint, metal-rich galaxies bright.  It is thus necessary to analyze hosts with a wide range of absolute magnitudes if we intend to measure a wide range of progenitor metallicities.  We demonstrate in Section~\ref{sec:res-complete} that the final sample we use in this paper, after all of our data cuts are applied, still satisfies this criterion, though we are left with a much reduced data set.

\subsection{Spectroscopic Data Processing}
The primary focus of this section is to describe the reduction of data obtained at the Gemini Observatory.  Information pertaining to the reduction of NTT spectra can be found in \citet{Ostman}, while the APO procedure is described in \citet{Zheng}.  A forthcoming paper \tolmstead\ will describe the data pipeline from BOSS.  All spectra from the SDSS spectroscopic survey used in this paper are from data release 7 \citep[DR7;][]{DR7}.

The faintest identified host galaxies of SNe Ia at $z < 0.15$ (determined from SN spectra) were observed with the Gemini Multi-Object Spectrograph (GMOS) at the Gemini-South observatory.  We were awarded 23 hours of time during semester 2008B (program GS-2008B-Q-38), during which 17 host spectra were obtained.  We took spectra in \textsc{longslit} mode with a 1.0 arcsec slit, no filter, and used the B600 grating.  The approximate observed wavelength coverage was 3800-6700 \AA.  Three exposures of duration 500-800 seconds each were taken for central wavelengths of both 520 and 525 nm, where the offset mitigates the effect of bad detector pixels in our reduced data.  The data were binned 2 times in the spatial direction and 2 times in the spectral direction for a wavelength dispersion of 0.9 \AA\ pixel$^{-1}$.  The B600 grating has a spectral resolution of $1688$ at its blaze wavelength of $\sim4610$\AA\ when used with a 0.5 arcsec slit.  

We performed most of the data reduction using version 1.10 of the Gemini IRAF\footnote{IRAF is distributed by the National Optical Astronomical Observatories, which are operated by the Association of Universities for Research in Astronomy, Inc., under cooperative agreement with the National Science Foundation.} package.  The raw two-dimensional (2D) spectra are in multi-extension FITS (MEF) format; we propagate the variance (VAR) and data quality (DQ) planes throughout our reduction.  We created mean bias frames by combining 5 bias frames from each observation night with {\tt gbias}, and normalized flat frames were generated using {\tt gsflat}.  All science images and our spectral standard (LTT9239) were then bias-subtracted, flat-fielded, and overscan subtracted with {\tt gsreduce}.  Wavelength calibration was performed using the task {\tt gswavelength} and the CuAr lamp spectra (also processed with {\tt gsreduce}) taken at the time of observation.  The wavelength calibration from the arc spectrum was applied to the 2D science spectra, which were then rectified using the task {\tt gstransform}.  The sensitivity function is determined by running the spectral standard through {\tt gsstandard}, which is then applied to the galaxy spectra with {\tt gscalibrate}.

We do not use {\tt gsskysub}, and instead apply {\tt gsextract} immediately.  This choice was made because {\tt gsextract} uses the routine {\tt apall}, which includes a sky subtraction procedure.  If one runs {\tt gsskysub} and then {\tt gsextract}, the resulting variance spectrum incorrectly contains low uncertainty for regions that originally featured heavy sky contamination, as the sky-lines have been removed in the previous step.  We also found that the Gemini tasks do not correctly propagate the VAR and DQ planes.  In {\tt gsextract}, the input DQ plane is not used, and the output DQ plane is simply set to be equal to 0 (i.e., `good') everywhere.  The VAR plane output is the square of the uncertainty spectrum determined by the call to {\tt apall}; the VAR plane that had been propagated up to this point is simply discarded.  We still use this spectrum to describe the uncertainties in our flux measurements as it is the same as we would have had, if we had performed the reduction procedure outside of the Gemini suite of packages.


\section{Emission-Line Analysis}
\label{sec:late}

Deriving the gas-phase oxygen abundance and the sSFR requires accurately measuring the flux in specific emission lines of the SNe Ia host galaxies; throughout this paper we will refer to the gas-phase oxygen abundance as `metallicity', unless the context requires more clarity.  Our procedure for obtaining emission-line spectra from observed data closely follows that in T04, in which a model of the continuum flux is subtracted from the observed spectrum.  This corrects for absorption features superimposed on emission lines.  We perform this analysis on all of our data, which totals 208 spectra from 138 host galaxies.

\begin{figure}[!htp]
  \centering
  \includegraphics[width=1.0\textwidth]{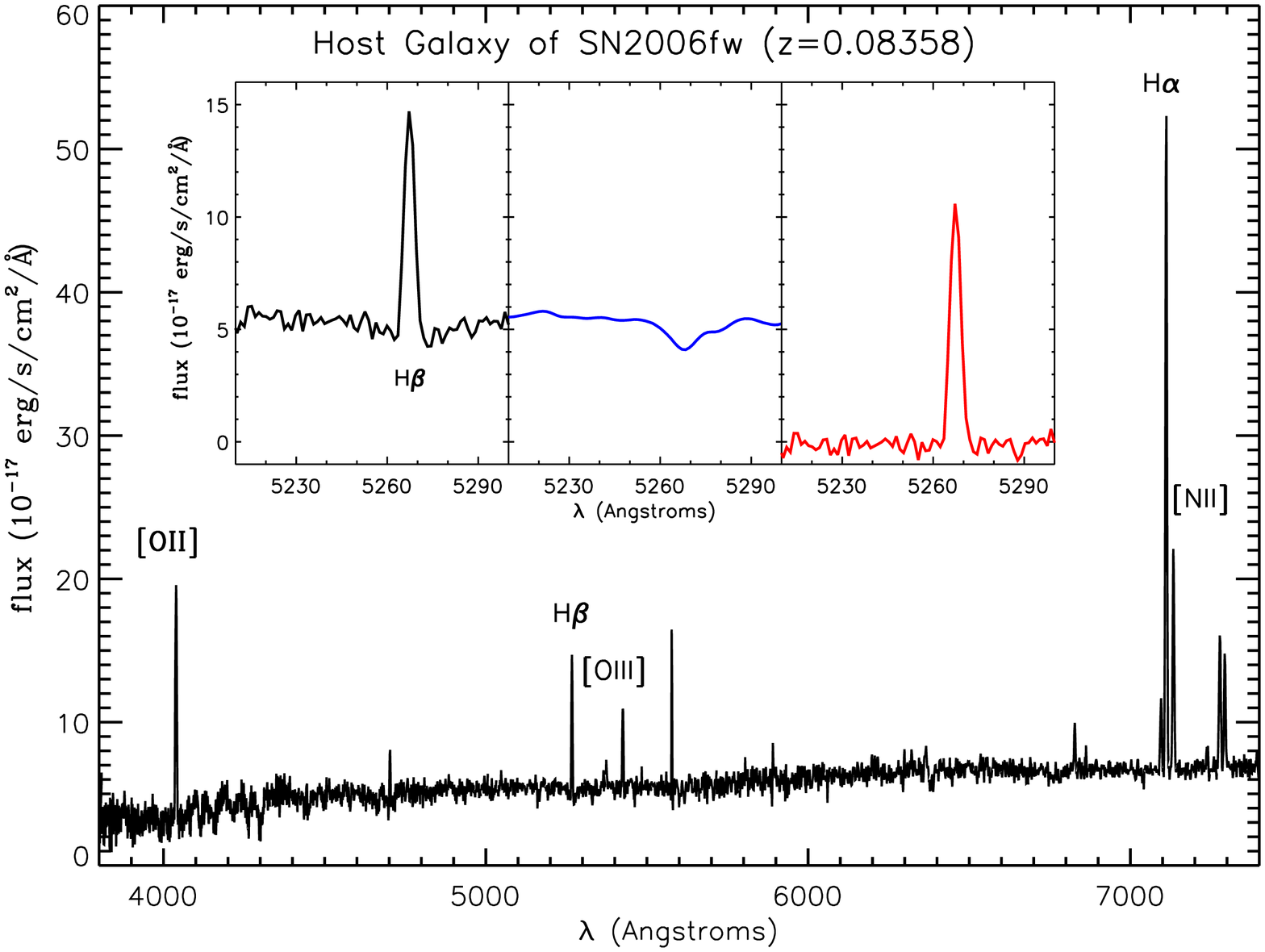}
  \caption{Spectrum of the host galaxy ($z=0.08358$, $m_r=18.4$) of SN 2006fw, obtained by BOSS \tolmstead.  Spectrum is shown in the observed frame, with Milky Way extinction correction from \citet{Schlegel}.  Key emission lines used throughout this paper are labelled.   The signal-to-noise ratio (S/N) of this spectrum at $\sim5000$\AA\ is $13$; note, however, that the S/N of the emission lines are typically much higher than that of the continuum.  INSET:  An example of the continuum subtraction procedure outlined in Section~\ref{sec:late}.  On the left is an enlargement of the full spectrum plotted below around the $H\beta$ line; the middle figure shows the best-fit model for the continuum; and the right shows the continuum-subtracted spectrum, which we refer to as the `emission-line' spectrum, and from which we measure fluxes.  
    \label{fig:spectra}}
\end{figure}

We begin\footnote{For spectra obtained with the SDSS 2.5m telescope, the wavelengths are first converted from their default values (`vacuum') to `air'.}  by correcting our spectra for extinction in the Milky Way along the line of sight \citep{Schlegel}, and then mask the spectra in the region about the Balmer lines (H$\alpha$, H$\beta$, H$\gamma$, H$\delta$) and the forbidden lines (\OII\ \dbll$3726, 3729$, \OIII\ \dbll$4959, 5007$, \NII\ \dbll$6548, 6584$, \SII\ \dbll$6717, 6731$).  We match this continuum spectrum with model spectral energy distributions (SEDs) drawn from a grid of Composite Stellar Populations (CSPs) by \citet[hereafter BC03]{BC03}, the parameters of which can be found in Table~\ref{tb:bc03}.  The model spectra are redshifted to the observed host-$z$, and the fluxes are convolved with a Gaussian to mimic both the instrumental dispersion and the host's intrinsic velocity dispersion (also listed in Table~\ref{tb:bc03}).  The best-fit CSP model from our grid is determined for the masked host spectrum through a least-squares analysis (see Figure~\ref{fig:spectra}).  The continuum-subtracted spectrum is then smoothed by subtracting off the average flux in a 200 pixel sliding window, with emission lines again masked from this average.  Fluxes are then measured by simulatenously fitting Gaussians to each of the aforementioned 12 lines, where the set of all Balmer lines and the set of all Forbidden lines are each constrained to have a single common line width and velocity offset, resulting in 16 free parameters for 12 lines.  Making this assumption strengthens the detection of weaker lines.

\begin{deluxetable}{l c}
\tablewidth{0pt}
\tabletypesize{\scriptsize}
\tablecaption{BC03 Composite Stellar Population Models\label{tb:bc03}}
\tablehead{
  \colhead{Parameter} &
  \colhead{Values} 
}
\startdata
IMF                                   & Chabrier \\
$[$Z/H$]$                             & 0.004, 0.008, 0.02, 0.05 \\
$\tau$ [Gyr]\tablenotemark{a}         & 0, 0.1, 1.0, 4.0 \\
Dust\tablenotemark{b}                 & 0, 0.25, 1.0, 2.0 \\
$\sigma_{\textrm{int}}$ [km s$^{-1}$] & 50, 100, 150, 200, 250, 300, \\ 
                                      & 350, 400 \\
Age [Gyr]                             & 0.1, 0.2, 0.3, 0.4, 0.5, 0.6, \\
                                      & 0.7, 0.8, 0.9, 1.0, 1.2, 1.4, \\
                                      & 1.6, 1.8, 2.0, 2.5, 3.0, 3.5, \\
                                      & 4.0, 4.5, 5.0, 6.0, 7.0, 8.0, \\
                                      & 9.0, 10.0, 11.0, 12.0, 13.0, 14.0 \\
\enddata
\tablenotetext{a}{Exponential Star Formation History, where SFR(t) $\sim e^{-t/\tau}$.  `0' denotes a single burst (SSP).}
\tablenotetext{b}{\emph{V}-band optical depth for stars younger then $10$ Myr.}
\end{deluxetable}

If the spectrum of a galaxy is dominated by an active galactic nucleus (AGN) instead of star formation, then emission-line ratios do not accurately reflect the gas-phase metallicity.  For this reason we use a BPT diagram (\OIII /H$\beta$ vs. \NII /H$\alpha$; \citealp{BPT}) to identify and remove galaxies with AGN contamination from our host sample.  We use the division in this plane between star-forming galaxies and AGN calibrated by \citet{Kewley01}, as well as the more conservative division by \citet{Kauffmann}; galaxies for which the two calibrations disagree are labelled `composite' \citep[similarly to][]{Brinchmann}.  We perform our subsequent analysis both with and without the `composite' hosts.  We adopt the T04 S/N requirements; the H$\alpha$, H$\beta$, and \NII\ $\lambda 6584$ lines are required to be detected at $>5\sigma$, and \OIII\ $>3\sigma$.   To avoid the removal of high-metallicity hosts (which have a large $\log{\textrm{\NII}/\textrm{\OIII}}$), spectra with \OIII\ $<3\sigma$ but $\log{(\textrm{H}\alpha/\textrm{\NII})} < -0.4$ are retained in our sample.

The remaining galaxy spectra are then corrected for host-galaxy extinction by assuming that the intrinsic H$\alpha$/H$\beta$ ratio follows Case-B Recombination \citep{Osterbrock}, and any deviations are due to extinction following the \citet{CCM} reddening curve, with $R_V=3.1$.  For the few cases where the observed H$\alpha$/H$\beta$ ratio is bluer than the assumed intrinsic value, we make the assumption that A$_V\approx 0$.

The spectra from Gemini are treated differently throughout this paper, as these only extend to $\approx 6600$\AA\ in the observed frame.  When the emission-line fluxes are fit we perform the same process as previously detailed, except we do not include constraints on H$\alpha$, \NII, or \SII\ (the Gemini hosts are all at $0.09 < z < 0.16$).  The lack of observed H$\alpha$ and \NII\ means we cannot use the BPT diagram to assess whether the spectrum is AGN or not.  While visual inspection shows that none of the emission lines are broad, AGN contamination cannot be completely ruled out.  For host-galaxy reddening corrections, the H$\gamma$/H$\beta$ ratio is used in place of H$\alpha$/H$\beta$.  We require H$\gamma$ to have S/N $>10$ for this correction, which holds true for most of our spectra.

We summarize the statistics of our sample in Table~\ref{tb:spec}.  We start with all 208 SN Ia host galaxy spectra, and then apply our cuts on the observed SN properties and on the host-galaxy spectral properties.  The total number of spectra that pass all of our cuts is greater than number of SN host galaxies in our final sample due to multiple observations of the same host.

\begin{deluxetable}{l cccc}
\tablewidth{0pt}
\tabletypesize{\scriptsize}
\tablecaption{Breakdown of Spectral Host Observations\label{tb:spec}}
\tablehead{
  \colhead{Telescope} &
  \colhead{Total} &
  \colhead{Light Curve} &
  \colhead{S/N} &
  \colhead{AGN\tablenotemark{a}} \\
  & Spectra & Cuts & Cuts & Cuts
}
\startdata 
SDSS    &  123  &  59  &  43  &  28/10 \\             
BOSS    &   54  &  31  &  21  &  17/2  \\
Gemini  &   16  &  12  &   9  &   9\tablenotemark{b}  \\
NTT     &    8  &   6  &   3  &   3/0  \\
APO     &    7  &   7  &   2  &   2/0  \\ 
\enddata
\tablecomments{~Light Curve cuts (Section~\ref{sec:data}) remove host-galaxy spectra from our sample for which the SN doesn't have a distance modulus measurement from \salt.  Signal-to-noise and AGN cuts (Section~\ref{sec:late}) remove spectra which cannot provide a metallicity or SFR measurement.  Numbers listed under the selection cuts are the quantity that passes each cut as well as cuts in the columns that precede it.}
\tablenotetext{a}{Spectra passing the AGN cuts are grouped as 'SFG/Composite'; see Section~\ref{sec:late}}
\tablenotetext{b}{As our observations with Gemini do not include $H\alpha$, we do not make any cuts for AGN activity.}
\end{deluxetable}

\subsection{Metallicity}
\label{sec:OH}

There exist many attempts at calibrating emission-line ratios to metallicity, both empirically derived and based on photoionization models \citep[for a summary, see][hereafter KE08]{KE08}.  While the absolute calibrations of models rarely agree, relative differences in metallicities are usually consistent - two different methods are likely to agree that one galaxy is more metal-rich than another.  However, the calibration offset precludes making measurements in one system and directly comparing them to a measurement made via a different method.  In KE08, fitting functions that transform metallicities computed in one system to those in another are determined for a variety of widely used emission-line techniques.  We use the diagnostic of \citet[hereafter KD02]{KD02} throughout our analysis, except in the case of our Gemini observations, for which we use \citet[hereafter KK04]{KK04}.

Galaxy metallicity is derived in KD02 through the ratio of \NII\ $\lambda$6584/\OII, which is shown in their Figure 3 to be insensitive to variations in the ionization parameter at super-solar oxygen abundances.  As discussed in Appendix A of KE08, for values of $\log{\textrm{\NII/\OII}} > -1.2$, the metallicity can be obtained as a root of the equation
\begin{align}
 \log({\textrm{\NII/\OII}}) &= 1106.8660 - 532.15451Z + 96.373260Z^2 \nonumber \\
                            & - 7.8106123Z^3 + 0.23928247Z^4,
\label{eq:KD02}
\end{align}
where $Z=\log([\textrm{O/H}])+12$.  The systematic uncertainty to these measurements is $\sim$0.1 dex (KE08).  For $\log{\textrm{\NII/\OII}<-1.2}$, the average of the $R_{23}$ calibrations by KK04 and \citet{M91} is used; this case applies to only 2 of our host galaxies.  

For the spectra taken at Gemini, we do not have observations of the \NII\ lines, so we must use an $R_{23}$-based method ($\textrm{R}_{23} \equiv (\textrm{\OII\ \dbll}3726,3729 + \textrm{\OIII \dbll}4959,5007)/\textrm{H}\beta$).  We choose to use the KK04 calibration for metallicity derivations of these galaxies, which has a systematic uncertainty of $\sim$0.15 dex (KE08).  The $R_{23}$ function is double-valued as a function of metallicity, with small values of $R_{23}$ being representative of both low and high oxygen abundances.  The KK04 method reaches a self-consistent solution for the ionization parameter and the metallicity by iteratively using the $R_{23}$ line ratio and the parameter $y \equiv \log\left({\textrm{\OIII}~\lambda 5007/\textrm{\OII}}\right)$, the latter of which is necessary for constraining the ionization state.  To break the degeneracy and determine to which of the two `branches' of the double-valued function a galaxy belongs, the \NII/\OII\ ratio is typically used.  Since \NII\ is not observed in our Gemini spectra, we convert the measured $R_{23}$ value for each galaxy into metallicities from both branches, and explore in Section~\ref{sec:res-metal} the implications from all combinations of these metallicities.  When we add the $R_{23}$-based metallicities from Gemini to the rest of our measurements, we use the KK04$\rightarrow$KD02 conversion function listed in Table 3 of KE08, which adds an RMS scatter of 0.05 dex (KE08).  As described there, the transformation function is only valid for KK04-based metallicities with values $8.2 < Z < 9.2$.  All emission lines used in this analysis are labelled in Figure~\ref{fig:spectra}. 

To truly measure the metallicity of the progenitor's environment, our spectra should be centered on the location of the SN.  However, all of the spectra used in this paper were \emph{not} obtained in this manner; both the fiber spectra and slit spectra were centered on the galaxy core.  The quantity that we are actually measuring, then, is the average metallicity from the integrated luminosity of a fraction of the host.  \citet{Kewley05} showed that for an emission-line metallicity measurement to be representative of the global value, the spectrum should contain $>20\%$ of the host-galaxy \emph{g}-band flux; KE08 showed that the \emph{g}-band flux fraction should be $>30\%$ for galaxies with $M > 10^{10} M_{\odot}$.  Thus, the fraction of the total host-galaxy light which is in our spectra is an important quantity to consider.  Spectra obtained as a part of the SDSS survey have fiber- and model-magnitudes associated with each galaxy, so we can easily compute the observed spectral fraction.  As the BOSS fiber spectrum is only sampling the central $2\arcsec$ in diameter, we cannot use the SDSS fiber magnitude, which was obtained with a $3\arcsec$ fiber.  The SDSS Catalog Archive Server (CAS) contains averaged surface brightness profiles within annuli of increasing radius.  We use these brightness profiles to obtain the total apparent magnitude within the central 3 annuli, which span a diameter of $\approx~2.05\arcsec$, and thus provides a good approximation to the BOSS fiber size.  For the Gemini observations, inspection of the finder images clearly showed that all observed host galaxies had a majority of their light contained within the slit width.  

\subsection{Star-Formation Rate}
\label{sec:SFR}

The recent star-formation rate of a galaxy can be estimated by the flux in the H$\alpha$ nebular line \citep{Kennicutt}, whose prominence is due to absorption and re-emission of stellar light bluewards of the Lyman Limit.  As only young, massive stars with $M > 10M_{\odot}$ contribute significant luminosity in this feature, the H$\alpha$ diagnostic is sensitive to star-formation in the past $\sim 10^7$years.  We adjust the conversion factor in \citet{Kennicutt} for a Chabrier IMF \citep{Chabrier} by dividing it by $1.7$ \citep{Pozzetti}, which gives

\begin{equation}
  {\rm SFR}~[{\rm M}_{\odot}~{\rm yr}^{-1}] = 10^{-41.33} ~L(H\alpha)~[{\rm erg~s}^{-1}]
  \label{eq:SFR}
\end{equation}

We conservatively assume a systematic uncertainty in $\log{[{\rm SFR}]}$ of $0.2$ based on \citet{Brinchmann}.  The $H\alpha$-luminosity ($L(H\alpha)$) is derived from the measured flux by assuming the best-fit cosmology detailed in Section~\ref{sec:data}.  Equation~\ref{eq:SFR} can only tell us the star-formation from the integrated light that comprises each spectrum; the SFR of the entire galaxy is also a function of the percentage of the total luminosity of the galaxy in each spectrum.  For galaxies with a spectrum from SDSS or BOSS, we estimate the total SFR for the galaxy by scaling the quantity derived from Equation~\ref{eq:SFR} by the inverse of the percentage of \emph{u}-band light contained in the fiber \citep[see][Appendix A]{Gilbank}.  This is obtained in a way analagous to that described for the \emph{g}-band light fraction in Section \ref{sec:OH}.  

We compute the sSFR of the host in two different ways.  In the first method, we divide the SFR obtained using Equation~\ref{eq:SFR} (the SFR within the fiber) by the mass derived from the least-squares fit of BC03 models to the continuum in Section~\ref{sec:late}.  Each model is normalized to the luminosity from 1${\rm M}_{\odot}$ of stars, so the multiplicative factor which minimizes the chi-squared of the fit for the data to each model (along with the distance modulus) provides a mass measurement.  For every galaxy we make a cumulative distribution function (CDF) of the mass probability.  We take the median mass value from this distribution as our estimate of the mass, with half the difference between the masses at 16\% and 84\% probability as our 1$\sigma$ error bar.  As the resulting sSFR is obtained directly from our spectroscopic observations, it is denoted as \ssfrs.  

The second method we use to derive the sSFR is to take the SFR scaled to the entire host galaxy and divide by the full mass of the galaxy, using the mass as determined from fits of the observed UV, optical, and near-IR photometry to galaxy SED models, done in \citet{Gupta}.  The best-fit mass in their work is similarly defined as being the median of the CDF from their model fits.  As this process requires the use of photometric measurements in obtaining both the mass and total SFR, we denote it \ssfrp.

\section{Results}
\label{sec:results}
In Table~\ref{tb:EZ} we list the derived oxygen abundances in 39 SN Ia host galaxies from useful spectra, excluding Gemini.  All hosts in this table have a metallicity computed directly from the \NII/\OII\ ratio via the KD02 method.  Specific star-formation rates from both methods detailed in the previous section are also listed.  Although there are more emission lines required at a specified S/N for the derivation of the metallicity than for the SFR, if the host cannot be placed on a BPT diagram then we are unable to determine whether the lines come from a star-forming region or AGN.  As such, we do not list any host spectra in Table~\ref{tb:EZ} with a SFR measurement and no metallicity.  Where there are multiple spectra from SDSS, we give the weighted average for each quantity.  We list measurements from different sources separately, as the derived quantities are dependent on the covering fraction of the spectrum.  Also included in Table~\ref{tb:EZ} are Hubble Residuals (HR) from SNANA using both \mlcs\ and \salt; the AGN contamination in the emission-line spectrum; and the fraction of the host's luminosity which is captured by the fiber, where applicable.

A total of 16 host galaxies at $z<0.15$ were observed with Gemini (one of the galaxies had $z=0.1543$).  We are able to calculate an $R_{23}$ value from the spectra in 11 of these galaxies, although two of these are host to SNe that fail the light-curve cuts for both \salt\ and \mlcs.  For the two host spectra with the largest $R_{23}$ values, the lower-branch metallicity derived from the fitting function of KK04 is higher than the upper-branch metallicity, indicating that we are unable to constrain the metallicity with this diagnostic.  In Table~\ref{tb:gem} we list the $R_{23}$ value; KK04 upper- and lower-branch metallicities; and the KD02-base metallicities for both branches, using the conversion factors in KE08.  These spectra are of strong emission-line galaxies, and as such our statistical uncertainties on $R_{23}$ and all quantities derived thereof are small; the systematic uncertainties from the KK04 method and from the KE08 conversion function dominate our quoted errors.  As in Table~\ref{tb:EZ}, Hubble Residuals and sSFR measurements are listed in Table~\ref{tb:gem}, including for hosts which do not have metallicity measurements.

In the analysis that follows we use the Markov Chain Monte Carlo (MCMC) package LINMIX \citep{Kelly07}, which allows for uncertainties in both the x- and y-coordinates, to linearly fit for Hubble Residuals as a function of host metallicity and sSFR.  We use this package because our dependent variables are derived quantities, subject not only to measurement uncertainties but systematic uncertainties from the functions which relate the observable to the desired physical property.  The linear regression coefficients and their significance are determined from the posterior distribution.  Following the work of \citet{Kelly10}, we cite both the percentage of runs from the posterior of the MCMC which have a slope with the opposite sign of the apparent correlation, and the significance at which the best-fit slope deviates from 0.

\begin{deluxetable}{l ccccccccr}
\tablewidth{0pt}
\tabletypesize{\scriptsize}
\tablecaption{Host Galaxy Metallicity and sSFR\label{tb:EZ}}
\tablehead{
  \colhead{IAU\tablenotemark{a}} &
  \colhead{Redshift} &
  \colhead{log[O/H]+12\tablenotemark{b}} &
  \colhead{HR$_{\textrm{MLCS}}$\tablenotemark{c}} &
  \colhead{HR$_{\textrm{SALT}}$\tablenotemark{c}} &
  \colhead{\ssfrs\tablenotemark{d}} &
  \colhead{\ssfrp\tablenotemark{d}} &
  \colhead{\emph{g}-band} &
  \colhead{AGN\tablenotemark{e}} &
  \colhead{Spectral} \\
  Name & & & & & & & fraction & & Source 
}
\startdata 
2005ez  &  0.1298  &  8.96(0.03)  &  -0.29(0.11)  &  -0.08(0.13)  &  -11.98  &  -11.52  &  0.52  &  1  &  SDSS \\
2005ff  &  0.0902  &  8.63(0.02)  &  -0.07(0.09)  &  -0.09(0.07)  &  -10.42  &   -9.77  &  0.19  &  1  &  BOSS \\
2005fv  &  0.1182  &  8.71(0.02)  &  -0.01(0.09)  &  -0.08(0.06)  &  -10.44  &   -9.91  &  0.21  &  0  &  SDSS \\
2005fw  &  0.1437  &  8.60(0.03)  &   0.18(0.07)  &   0.04(0.06)  &   -9.75  &    ---   &  ---   &  0  &  NTT  \\
2005gp  &  0.1266  &  8.92(0.02)  &  -0.20(0.12)  &  -0.23(0.07)  &  -10.56  &   -9.75  &  0.29  &  0  &  SDSS \\
2005hn  &  0.1076  &  8.85(0.02)  &   0.09(0.11)  &   0.03(0.07)  &   -9.60  &   -9.77  &  ---   &  0  &  APO  \\
        &          &  8.91(0.01)  &               &               &  -10.16  &          &  0.47  &  0  &  BOSS \\
2005gb  &  0.0866  &  8.95(0.01)  &   0.16(0.07)  &  -0.04(0.06)  &  -10.37  &   -9.60  &  0.17  &  1  &  SDSS \\
2005ho  &  0.0628  &  8.62(0.01)  &   0.04(0.08)  &  -0.08(0.08)  &   -9.81  &   -9.26  &  0.22  &  0  &  SDSS \\
2005hx  &  0.1210  &   ---\tablenotemark{f} & 0.14(0.12) & 0.02(0.15) & -9.96 &  -9.05  &  0.29  &  0  &  BOSS \\
2005if  &  0.0671  &  8.92(0.01)  &   0.01(0.09)  &  -0.12(0.09)  &  -10.22  &   -9.60  &  0.16  &  0  &  SDSS \\
6213    &  0.1094  &  9.03(0.01)  &   ---         &   0.24(0.21)  &  -10.39  &    ---   &  0.09  &  0  &  SDSS \\
2005ij  &  0.1246  &  8.83(0.01)  &  -0.08(0.07)  &  -0.12(0.05)  &  -10.59  &   -9.85  &  0.20  &  0  &  SDSS \\
2005ir  &  0.0764  &  9.00(0.01)  &   0.09(0.11)  &  -0.04(0.07)  &  -10.61  &   -9.71  &  0.14  &  0  &  SDSS \\
2005kp  &  0.1178  &  8.25(0.09)  &   0.12(0.13)  &   0.07(0.07)  &   -9.50  &   -8.85  &  0.34  &  0  &  BOSS \\
        &          &  8.37(0.03)  &               &               &   -9.53  &          &  ---   &  0  &  NTT  \\
2005ld  &  0.1453  &  8.72(0.01)  &   0.08(0.12)  &   ---         &   -9.80  &   -9.55  &  0.22  &  0  &  BOSS \\
2006fc  &  0.1217  &  8.76(0.01)  &   ---         &   0.00(0.07)  &  -10.21  &   -9.53  &  0.19  &  0  &  BOSS \\ 
2006fw  &  0.0835  &  8.81(0.01)  &   0.12(0.08)  &   0.01(0.06)  &  -10.19  &   -9.42  &  0.37  &  0  &  BOSS \\
2006fy  &  0.0827  &  8.79(0.01)  &   0.26(0.06)  &   0.03(0.06)  &  -10.10  &   -9.63  &  0.30  &  0  &  SDSS \\
2006fm  &  0.1257  &  8.85(0.01)  &   0.12(0.08)  &   0.06(0.06)  &  -10.64  &  -10.10  &  0.10  &  0  &  BOSS \\
2006hl  &  0.1482  &  8.85(0.01)  &  -0.04(0.08)  &  -0.10(0.06)  &  -10.60  &   -9.98  &  0.15  &  0  &  BOSS \\
2006hx  &  0.0454  &  9.10(0.02)  &  -0.26(0.11)  &  -0.22(0.10)  &  -11.18  &  -10.44  &  0.22  &  1  &  SDSS \\
2006kd  &  0.1363  &  9.06(0.02)  &   0.29(0.08)  &   0.28(0.06)  &  -10.69  &   -9.90  &  0.29  &  0  &  SDSS \\
        &          &  9.03(0.01)  &               &               &  -10.69  &          &  0.17  &  0  &  BOSS \\
15362   &  0.1341  &  8.64(0.03)  &  -0.24(0.23)  &  -0.02(0.27)  &  -10.14  &    ---   &  ---   &  0  &  APO  \\
        &          &  8.86(0.01)  &               &               &  -10.33  &          &  0.57  &  0  &  BOSS \\
2006la  &  0.1267  &  8.44(0.02)  &   0.49(0.09)  &   0.45(0.06)  &   -9.42  &   -9.64  &  0.53  &  0  &  BOSS \\
2006nc  &  0.1240  &  8.89(0.01)  &  -0.03(0.13)  &  -0.03(0.15)  &  -10.79  &   -9.74  &  0.18  &  0  &  BOSS \\
2006nd  &  0.1288  &  8.95(0.01)  &  -0.07(0.10)  &  -0.07(0.06)  &  -10.03  &   -9.43  &  0.28  &  0  &  SDSS \\
2007hx  &  0.0798  &  8.50(0.01)  &   0.32(0.10)  &   0.20(0.08)  &  -10.43  &  -10.01  &  0.07  &  1  &  SDSS \\
        &          &  8.61(0.02)  &               &               &  -10.11  &          &  0.05  &  0  &  BOSS \\
2007jt  &  0.1448  &  8.96(0.01)  &   0.04(0.08)  &  -0.10(0.05)  &  -10.18  &   -9.82  &  0.23  &  0  &  SDSS \\
        &          &  9.01(0.01)  &               &               &  -10.10  &          &  0.14  &  0  &  BOSS \\
2007ju  &  0.0636  &  8.26(0.03)  &   0.04(0.10)  &   0.02(0.10)  &  -10.33  &   -9.73  &  0.20  &  0  &  BOSS \\
2007jg  &  0.0371  &  8.79(0.01)  &   0.35(0.13)  &   0.16(0.12)  &  -10.69  &   -9.75  &  0.10  &  0  &  SDSS \\
2007jd  &  0.0727  &  9.05(0.01)  &  -0.01(0.13)  &  -0.03(0.08)  &  -10.68  &   -9.94  &  0.16  &  0  &  SDSS \\  
2007lg  &  0.1104  &  9.08(0.01)  &  -0.17(0.08)  &  -0.18(0.06)  &  -11.12  &  -10.75  &  0.42  &  0  &  BOSS \\
2007lo  &  0.1384  &  8.53(0.04)  &   0.12(0.10)  &   0.05(0.06)  &  -10.26  &   -9.76  &  0.42  &  0  &  BOSS \\
2007lc  &  0.1150  &  8.96(0.01)  &  -0.04(0.09)  &  -0.19(0.07)  &  -10.75  &  -10.03  &  0.23  &  0  &  SDSS \\
2007ma  &  0.1073  &  8.94(0.01)  &   0.03(0.08)  &   ---         &  -10.17  &   -9.50  &  0.29  &  0  &  SDSS \\
2007mh  &  0.1278  &  9.07(0.01)  &   0.14(0.08)  &   0.05(0.06)  &  -10.70  &  -10.02  &  0.27  &  1  &  SDSS \\
19048   &  0.1368  &  9.07(0.01)  &   0.78(0.16)  &   0.71(0.13)  &  -10.26  &    ---   &  0.53  &  0  &  SDSS \\
        &          &  9.08(0.01)  &               &               &  -10.39  &          &  0.43  &  1  &  BOSS \\
2007mn  &  0.0769  &  8.83(0.01)  &   0.20(0.07)  &   0.14(0.06)  &  -11.06  &  -10.43  &  0.16  &  1  &  SDSS \\
2007ou  &  0.1132  &  8.91(0.01)  &   0.48(0.10)  &   0.45(0.08)  &   -9.99  &   -9.44  &  0.23  &  0  &  SDSS \\
2007pd  &  0.1399  &  8.63(0.01)  &   0.10(0.10)  &   0.11(0.07)  &   -9.95  &   -9.97  &  0.18  &  0  &  BOSS \\
\enddata
\tablenotetext{a}{SNe in our Phot Ia sample do not have an IAU designation, and instead the internal SDSS-Supernova Survey candidate ID is given.}
\tablenotetext{b}{Metallicty values listed here are in units of 12+$\log{\textrm{[O/H]}}$.  For host galaxies where there is more than one spectrum from SDSS-I, we give a weighted average of our derived metallicities.  Errors listed here are statistical uncertainties; a systematic uncertainty of 0.1 dex is added in quadrature for all analyses.} 
\tablenotetext{c}{An additional uncertainty of 0.14 magnitudes is added in quadrature to the listed uncertainty in all analyses \citep[e.g., ][]{Kelly10}; this quantity is the intrinsic uncertainty in the SN Ia sample, required for the reduced $\chi^2$ of the best-fit cosmology to be $\approx$ 1.}
\tablenotetext{d}{$\log{\textrm{sSFR} [\textrm{yr}^{-1}]}$, derived from H$\alpha$ flux.  We adopt a systematic uncertainty of 0.2 dex for all SFR measurements.}
\tablenotetext{e}{`0' denotes a Star-Forming Galaxy (SFG), `1' a Composite.  Both are classified as star-forming by the criteria of \citet{Kewley01}, but the latter are classified as AGN by that of \citet{Kauffmann}}
\tablenotetext{f}{2005hx has $\log{\textrm{[N~II]/[O~II]}<-1.2}$, but the $R_{23}$ value is too high to constrain metallicity, as is also the case for two of our Gemini-based spectra.}
\end{deluxetable}

\begin{deluxetable}{l cccccccc}
\tablewidth{0pt}
\tabletypesize{\scriptsize}
\tablecaption{Host Galaxy Metallicity and sSFR from Gemini Observations\label{tb:gem}}
\tablehead{
  \colhead{IAU Name} &
  \colhead{Redshift\tablenotemark{a}} &
  \colhead{log($R_{23}$)} &
  \colhead{Z$_{\textrm{KK04}}$\tablenotemark{b}} &
  \colhead{Z$_{\textrm{KD02}}$\tablenotemark{c}} &
  \colhead{Z$_{\textrm{KD02}}$} &
  \colhead{HR$_{\textrm{MLCS}}$} &
  \colhead{HR$_{\textrm{SALT}}$} &
  \colhead{\ssfrs\tablenotemark{d}} \\
  & & & & lower branch & upper branch & & &
}
\startdata 
2005hr  &  0.1163  &  0.908(0.006)  &  8.45/8.51  &  8.34(0.15)  &  8.39(0.15) &  0.17(0.07)  &  0.13(0.06)  &  -9.22  \\
2005hx  &  0.1210  &  0.763(0.007)  &  8.23/8.77  &  8.14(0.13)  &  8.66(0.17) &  0.14(0.13)  &  0.02(0.15)  &  -9.96  \\
2006jh  &  0.1249  &  0.737(0.003)  &  8.13/8.81  &  ---         &  8.72(0.17) &  0.21(0.08)  &  0.17(0.06)  &  -9.82  \\ 
2006iz  &  0.1363  &  1.021(0.005)  &  ---        &  ---         &  ---        &  0.23(0.07)  &  0.21(0.10)  &  -9.24  \\
2006jq  &  0.1276  &  0.695(0.006)  &  8.21/8.84  &  8.13(0.13)  &  8.74(0.17) &  0.11(0.07)  &  0.04(0.05)  &  -9.92  \\
2006kb  &  0.1392  &  0.810(0.009)  &  8.35/8.68  &  8.25(0.14)  &  8.57(0.16) &  0.21(0.11)  &  0.07(0.13)  &  -9.68  \\
2006la  &  0.1270  &  0.903(0.005)  &  8.47/8.51  &  8.36(0.15)  &  8.39(0.15) &  0.49(0.09)  &  0.45(0.06)  &  -9.42  \\ 
2007lo  &  0.1386  &  0.679(0.017)  &  8.25/8.85  &  8.16(0.14)  &  8.76(0.18) &  0.12(0.10)  &  0.05(0.06)  &  -9.95  \\
2007lt  &  0.1140  &  0.971(0.010)  &  ---        &  ---         &  ---        &  0.30(0.13)  &  0.23(0.15)  &  -8.86  \\
\enddata
\tablecomments{Metallicity values listed here are in units of 12+$\log{\textrm{[O/H]}}$}
\tablenotetext{a}{All Hosts listed here have a statistical uncertainty in redshift of 0.0005}
\tablenotetext{b}{Metallicities derived assuming the lower/upper branches of the $R_{23}$ relation.}
\tablenotetext{c}{The lower limit of applicability of the KE08 transformation from KK04 to KD02 is \\ 12+log[O/H]$_{\textrm{KK04}}$ = 8.2, which gives an upper limit to any lower value, in the KD02 system, of 8.11.}
\tablenotetext{d}{$\log{\textrm{sSFR} [\textrm{yr}^{-1}]}$, derived from H$\beta$ flux; we assume that the intrinsic H$\alpha$/H$\beta$ ratio follows Case-B Recombination \citep{Osterbrock}.  We adopt a systematic uncertainty of 0.2 dex for all SFR measurements.}
\end{deluxetable}

\subsection{HR vs. Metallicity}
\label{sec:res-metal}
\begin{figure}[!htp]
  \centering
  \includegraphics[width=1.0\textwidth]{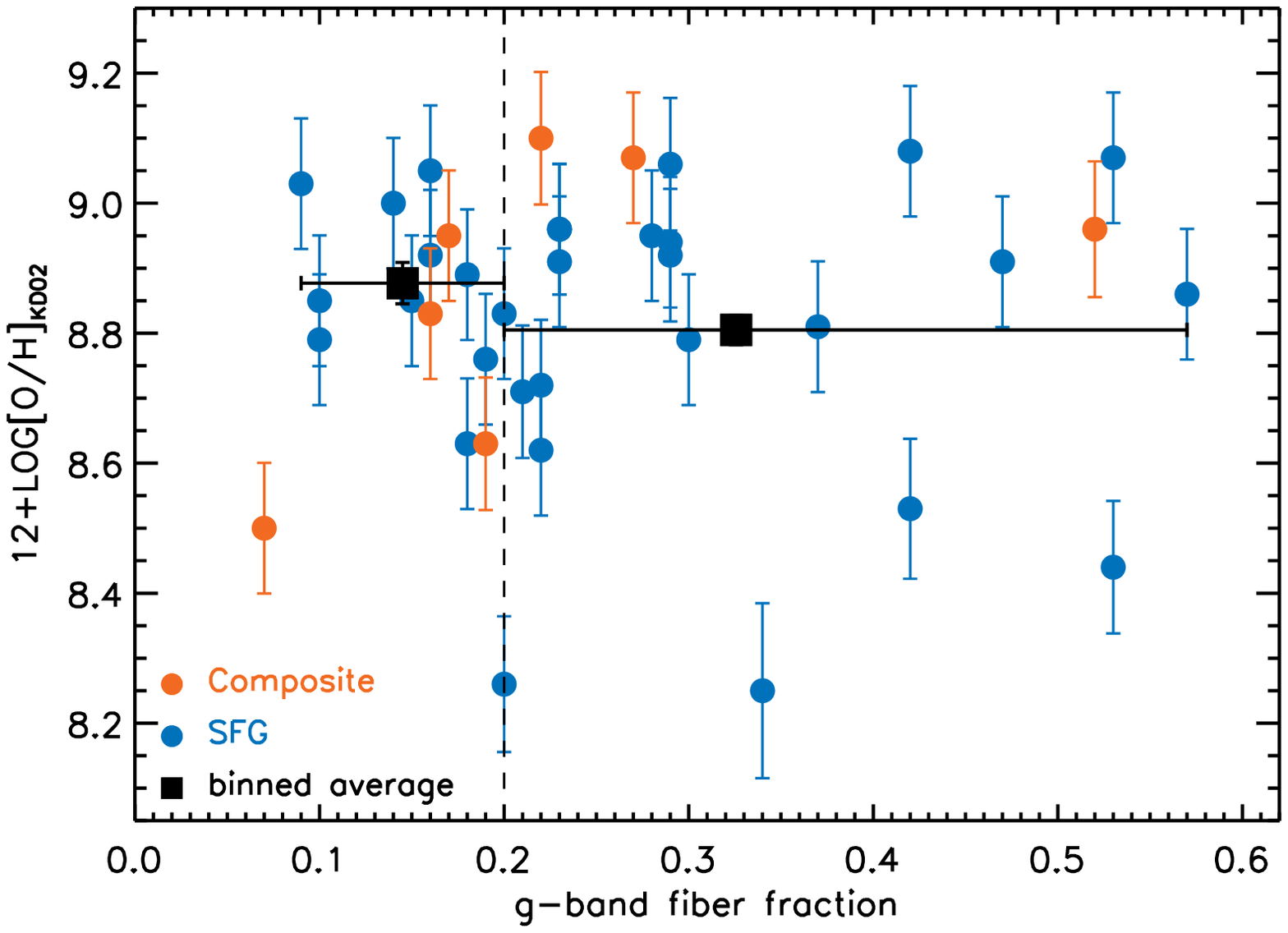}
  \caption{Metallicity measurements from SDSS-I and BOSS spectra as a function of host \emph{g}-band luminosity captured within the fiber.  Blue circles denote SFG, Red circles are Composites.  The dashed line shows the division employed by \citet{Kewley05} to distinguish between spectra which represent a global average and those which do not.  As expected, the average metallicity in low fiber-fraction spectra of SFG is higher than that in higher fiber-fraction spectra (black squares), though the difference is small ($0.072\pm0.039$ dex).  If the `Composite' spectra are included, the average is the same to better than 0.01 dex.
    \label{fig:metal-fiber}}
\end{figure}

The \emph{g}-band fiber fractions for BOSS and SDSS spectra are listed in Table~\ref{tb:EZ}, which we use together with the guideline of \citet{Kewley05} to determine whether a given fiber spectrum is representative of a `global' metallicity.  However, supposing that the metallicity of the progenitor is correlated with the globally averaged value, a measurement from a region that is simply biased with respect to the average would still be useful.  Including these measurements could add additional noise to our results, although this option would make our SFG sample almost $50\%$ larger.  In Figure~\ref{fig:metal-fiber} we examine the metallicity measurements from low fiber-fraction spectra.  For SFG, these spectra on average measure slightly more metal-rich regions with a smaller dispersion than high fiber-fraction spectra, although this distinction disappears when `Composites' are included.  The low fiber-fraction sample does not include any extremely high metallicities when compared to the higher fiber-fraction sample.  While we conclude that it is beneficial to use our full data set, for clarity we will quote our results from both a `global' sample and `full' sample in this section, where the former refers to \emph{g}-band fraction $\ge 0.20$, and the latter sample includes all SDSS and BOSS measurements.  For the slit-spectra in our data (NTT, APO, Gemini) the majority of the host was included in each observation, and these are included in both of our samples.  

For our observations using Gemini, we have upper- and lower-branch metallicity measurements, but lack the necessary data to discern the correct value.  We resolve this issue by determining the best-fit linear relation between HRs and metallicity in a given sample for each possible combination of Gemini-observed host metallicities.  The mean value of the significance over the whole set of possible metallicity combinations is taken, and we give limits based on the combinations which yield the lowest and highest significance.  
 
\begin{figure}[!htp]
  \centering
  \includegraphics[width=1.0\textwidth]{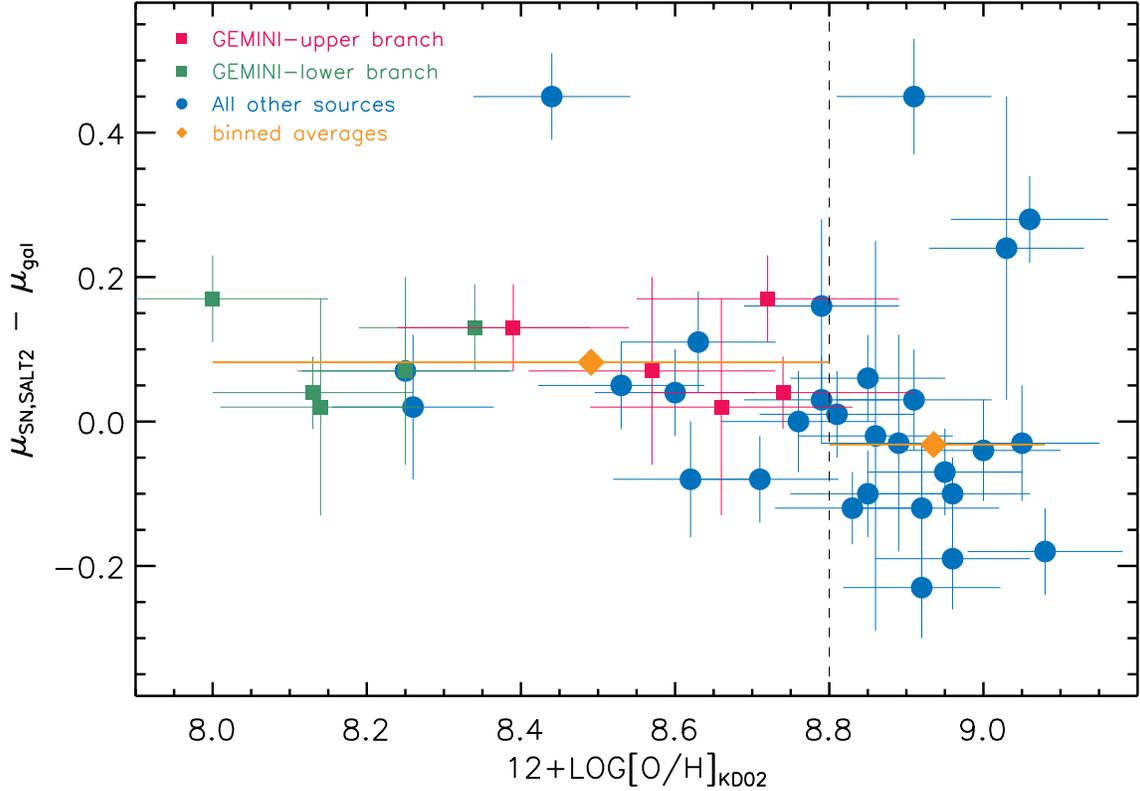}
  \caption{Hubble Residuals as a function of emission-line metallicity, measured using the KD02 method described in Section \ref{sec:late}.  Blue circles are SFG in our `full' sample, excluding Gemini sources.  The lower- and upper-branch metallicities from Gemini spectra are green and red squares, respectively.  The plotted error bars include the systematic uncertainty of 0.1 dex added in quadrature to log[O/H] measurements (KE08).  The orange diamonds represent the weighted averages of the upper- and lower-metallicity bins, with the division at $12+\log[{\rm O/H}]=8.80$ set so that the bins contain approximately the same amount of data (18 and 16 hosts, respectively).  We do not fold into our HR uncertainties the `intrinsic scatter' of $\approx$~0.14 mag used to bring the reduced $\chi^2$ of the best-fit cosmology down to 1; see Section~\ref{sec:res-metal} for a discussion. 
    \label{fig:metal}}
\end{figure}

A correlation between metallicity and Hubble Residuals is measured with LINMIX at the $1.3-2.0\sigma$ confidence level, depending on the combination of light-curve fitter and data sample used (see Table~\ref{tb:metal}).  The trend is in the direction where lower metallicities yield underluminous SNe after light-curve corrections, as inferred in \citet{Sullivan10} and predicted in \citet{KRW}.  We show the \salt\ HR vs. metallicity plot for the `full' sample of SFG in Figure~\ref{fig:metal}.  The hosts targeted by Gemini, being the faintest hosts within the redshift limit of our sample, unsurprisingly provide most of the weight in the low-metallicity region.  The plot excludes the metallicity measurements from 2 of the 7 Gemini-observed host galaxies:  those of SN 2006la and SN 2007lo.  Since both of these hosts also have BOSS spectra, which include direct observation of the H$\alpha$ line, we used the derived quantities for these hosts in place of those taken from Gemini spectra.  We note that the lower- and upper-branch Gemini metallicities of SN 2006la (8.36/8.39) agree with the BOSS value (8.44), and that the SN 2007lo values, while different (8.16/8.79 for Gemini, 8.53 for BOSS), both suggest a lower metallicity than the median from our sample.

The low-metallicity side of Figure~\ref{fig:metal} contains almost exclusively hosts with positive HRs.  We divide our samples into two metallicity groups about the line $\log[\textrm{O/H}]+12 = 8.80$, and compute the weighted average Hubble Residual and the error in the mean for each bin.  We choose this value for our metallicity division over a more natural choice of solar-metallicity ($\log[\textrm{O/H}]+12 = 8.69$; \citealp{Asplund}) for two reasons.  First, our value better splits the two samples into roughly equal sized bins, yielding 16 (low-Z) and 18 (high-Z) SFG for the `full' sample, and 13 (low-Z) and 11 (high-Z) SFG for the `global' sample.  Second, we can only include Gemini-based measurements if both possible $R_{23}$-based metallicities fall within the same bin; a solar-metallicity division would leave out two host galaxies whose branches straddle the dividing line, while our choice retains all Gemini measurements.  As can be seen in Figure~\ref{fig:metal}, the HRs differ significantly between the two bins, with a weighted-average of $-0.032\pm0.016$ in the high-metallicity bin and $0.082\pm0.017$ in the low-metallicity bin.   There is a 4.9$\sigma$ significance ($0.114\pm0.023$) in the difference between the mean \salt\ HR in SFG hosts in our `full' sample between low- and high-metallicity environments.  Similar significance is found if the smaller `global' sample is used in place of the `full' ($0.116\pm0.027$); if Composites are added to the SFG ($0.091\pm0.021$); or if \mlcs\ is used instead of \salt\ ($0.132\pm0.031$). 

In contrast with previous studies \citep{Kelly10,Sullivan10,Lampeitl10}, we do not include the intrinsic scatter in the Hubble Diagram (here $\approx0.14$ mag) in our HR uncertainties.  The intrinsic scatter is the quantity which must be added into the uncertainties of the measured distance moduli so that the reduced $\chi^2$ of the best-fit cosmology equals one.  As such, it represents a physical process unaccounted for in our light-curve fitters that affects the overall luminosity of the SNe Ia.  When attempting to explore the origin of this scatter, it is the measured difference between the distance modulus from each SN Ia, derived from known physical relations, and the distance modulus at its redshift for a given cosmology, that are of interest.  Adding in an extra source of uncertainty in the measured Hubble Residual is equivalent to diluting the confidence in which the light-curve fitter is returning an over- or under-luminous SN.  

\begin{deluxetable}{l cc}
\tablewidth{0pt}
\tabletypesize{\scriptsize}
\tablecaption{Significance of HR correlations with Gas-Phase Metallicity\label{tb:metal}}
\tablehead{
  \colhead{Sample} &
  \colhead{\salt} & 
  \colhead{\mlcs} 
}
\startdata 
`Global':SFG             &  7.9\%(1.3$^{+0.3}_{-0.2}\sigma$)  &  5.8\%(1.5$^{+0.3}_{-0.3}\sigma$)  \\
`Global':SFG+Composites  &  8.4\%(1.3$^{+0.3}_{-0.2}\sigma$)  &  5.7\%(1.5$^{+0.3}_{-0.3}\sigma$)  \\
`Full':SFG               &  2.0\%(1.9$^{+0.3}_{-0.2}\sigma$)  &  2.8\%(1.8$^{+0.3}_{-0.2}\sigma$)  \\
`Full':SFG+Composites    &  2.0\%(2.0$^{+0.2}_{-0.3}\sigma$)  &  4.0\%(1.7$^{+0.3}_{-0.3}\sigma$)  \\
\enddata
\tablecomments{Entries are the percent of slopes in the posterior distribution of LINMIX with sign opposite that of the best-fit value, with the mean significance of the deviation from 0, based on all possible combinations of Gemini-based metallicities, in parentheses.  The upper and lower limits on this significance reflect the strongest and weakest correlations for any single Gemini metallicity combination.  Note that the significance of these slopes is lower in all cases from the significance at which HRs in the lower-metallicity bin differ from those in the higher-metallicity bin (see Section~\ref{sec:res-metal}).}
\end{deluxetable}

\subsection{HR vs. sSFR}
\label{sec:res-ssfr}
\begin{figure}[!htp]
  \centering
  \includegraphics[width=1.0\textwidth]{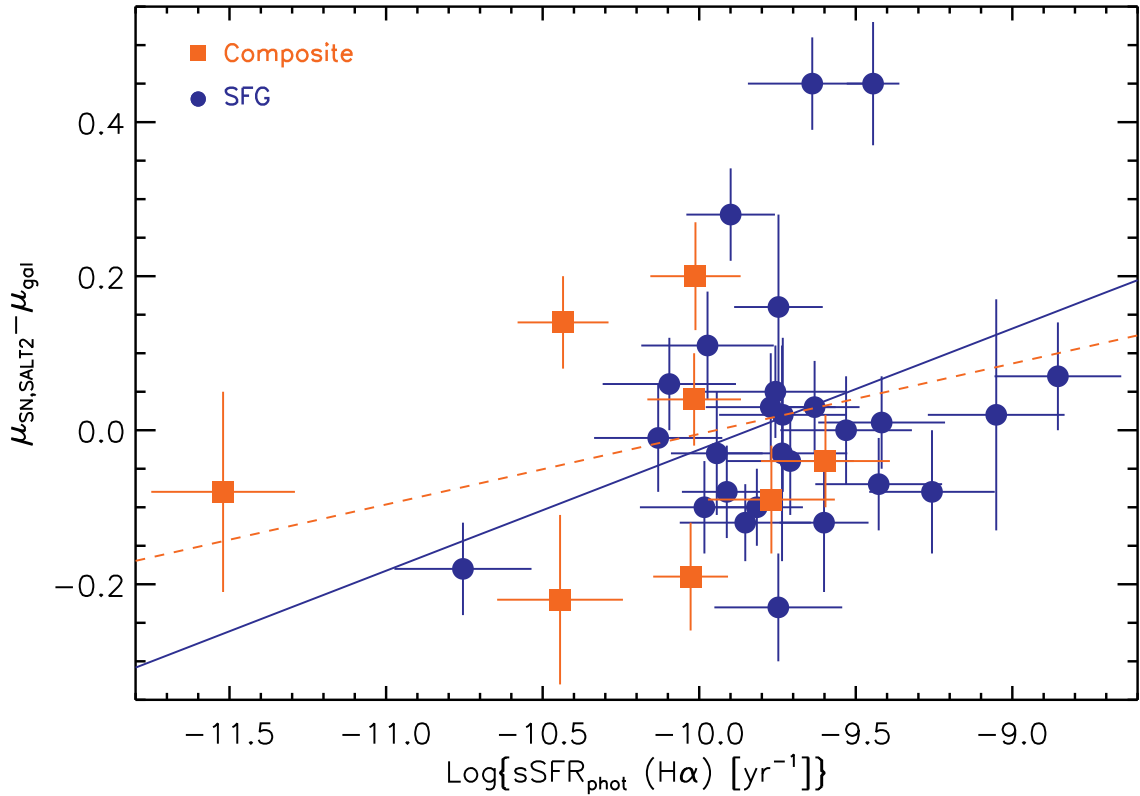}
  \caption{\salt\ HRs as a function of the \emph{photometrically} defined specific star-formation rate, \ssfrp.  Blue circles are emission-line hosts classified as star-forming galaxies (SFGs); red squares are Composites.  Neither the best-fit line for the SFG-only sample (solid line) nor the SFG+Composite sample (dashed line) are good fits to the data, though they hint at a general trend of overluminous SNe being preferentially found in galaxies with a low sSFR.  
    \label{fig:ssfrp}}
\end{figure}

We examined the correlation in actively SFG of both galaxy-averaged quantities, \sfrp\ and \ssfrp, with Hubble Residuals.  We find no correlation between HR and \sfrp; the best-fit slope deviates from zero at $\leq 1\sigma$ for both the SFG-only sample and the SFG+Composite sample (as defined in Section~\ref{sec:late}), whether we use \mlcs\ or \salt\ (See Table~\ref{tb:sfr}).  This result is not unexpected; a given SFR could denote either a massive galaxy with a low sSFR or a small galaxy with a high sSFR.  Thus it seems likely that a global SFR measurement, like the one used here, would show no correlation with HR.  Turning our attention to the sSFR, we use galaxy masses from \citet{Gupta} to compute the HR--\ssfrp\ relation, shown in Figure~\ref{fig:ssfrp}.  The evidence for a correlation using LINMIX is slightly more significant here than for \sfrp\, but the relation is poorly described by our linear fit, as the derived sSFRs display both a tight distribution and high scatter.  

\begin{figure}[!htp]
  \centering
  \includegraphics[width=1.0\textwidth]{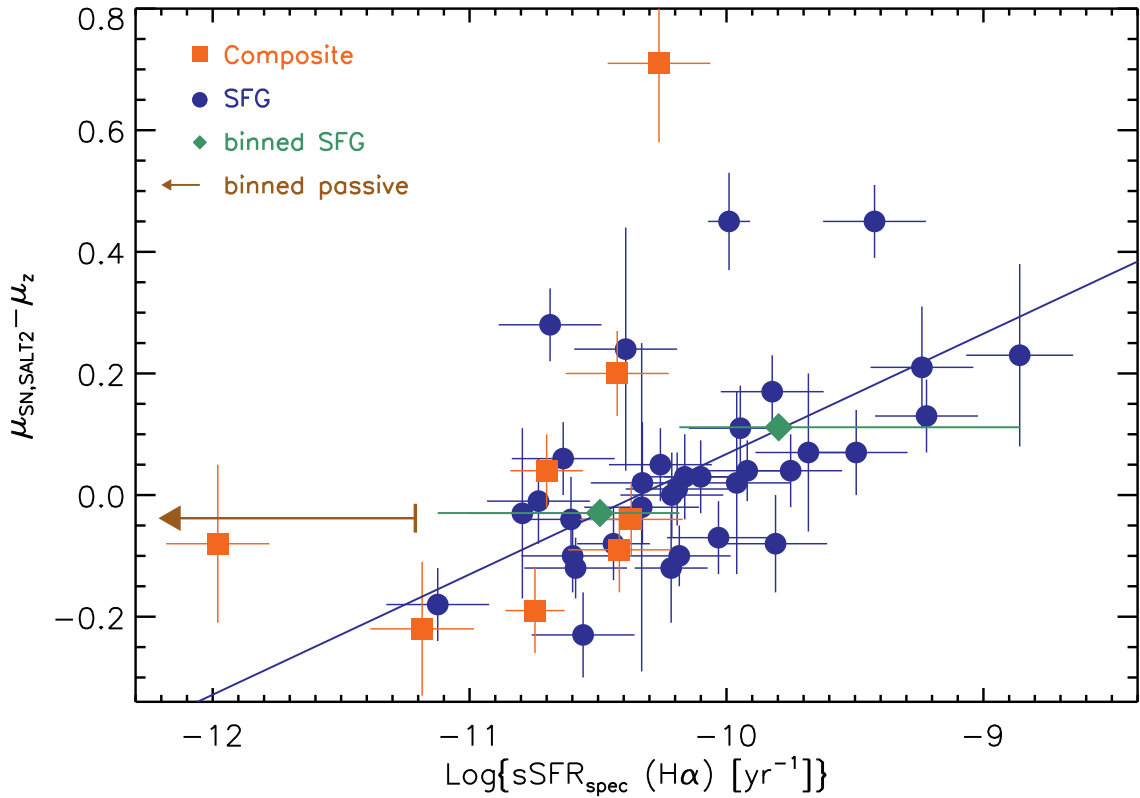}
  \caption{\salt\ HRs as a function of the \emph{spectroscopically} defined specific star-formation rate, \ssfrs.  Blue circles are hosts classified as star-forming galaxies (SFGs); red squares are Composites.  The best-fit line for the SFG-only sample is shown.  SNe Ia which are underluminous after light-curve corrections are preferentially found in regions of high sSFR.  If divided into equally populated bins (green diamonds), the bin of lower sSFR has a typical HR consistent with that of passive hosts (brown arrow), the upper limit of which is show here.
    \label{fig:ssfrs}}
\end{figure}

The results of our study of the correlations between the \ssfrs\ and HR are also found in Table~\ref{tb:sfr}.  As before, we show our results using both \mlcs\ and \salt\ HRs, and split our samples into SFG-only and SFG+Composite.  We show in Figure~\ref{fig:ssfrs} the \salt\ HRs as a function of our derived \ssfrs\ values.  We find a correlation between HR and \ssfrs\ at a confidence greater than 3$\sigma$ irrespective of our choice of light-curve fitter, in the sense that overluminous SNe Ia after corrections tend to reside in hosts with lower sSFR.  It is worth noting that only a fraction of the total intrinsic scatter in SN Ia luminosity is explained by this correlation.  Subtracting the best-fit line in Figure~\ref{fig:ssfrs} from the HRs of the SFGs, the intrinsic scatter is reduced from $0.14$ to $0.11$ mag.

\begin{figure}[!htp]
  \centering
  \includegraphics[width=1.0\textwidth]{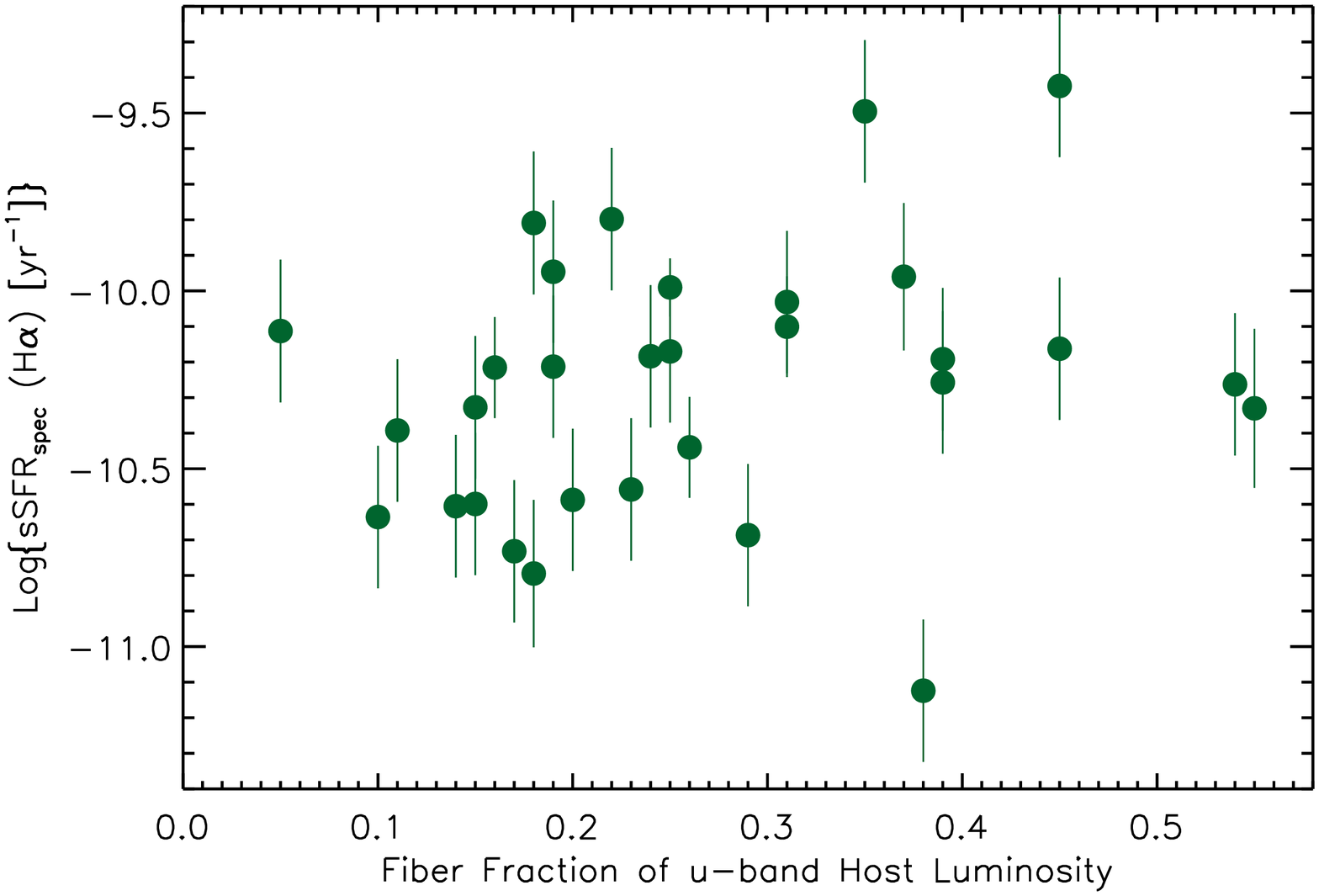}
  \caption{Specific star-formation rate measured within an SDSS-I or BOSS fiber, as described in Section \ref{sec:SFR}, as a function of the fraction of total \emph{u}-band flux from the galaxy captured within the fiber.  The absence of a correlation between \ssfrs\ and fiber-fraction suggests that an apparent dependence of Hubble Residuals on \ssfrs\ is not a function of our measurements originating from different physical scales.
    \label{fig:uband}}
\end{figure}

In the case of \ssfrp, the property that we are computing -- a galaxy-averaged sSFR -- should be directly comparable from one host to another.  However, the \ssfrs\ measurements are based on observations that measure different proportions of the host's total luminosity, and as such are not obviously directly comparable.   To explore what type of systematic effect this might create, we show in Figure~\ref{fig:uband} our derived \ssfrs\ as a function of \emph{u}-band fiber fraction.  It is clear from this figure that our measured \ssfrs\ is not simply a function of the fixed physical fiber size from SDSS and BOSS.  Thus the relationship we find between \ssfrs\ and HR cannot be dismissed as an artifact of the varying physical scales upon which our measurements are based.  We do not correlate our measured \sfrs\ with HR because the measured SFR, being a cumulative quantity, will depend strongly on the percentage of a galaxy's physical size enclosed in a given observation; this is why we previously chose to examine the global quantity \sfrp.

Our results for \ssfrs\ presented in Figure~\ref{fig:ssfrs} show a rapidly evolving relationship in sSFR, while \ssfrp\ does not.  This result lends itself to two questions.  First, why do the two differ?  The \ssfrp\ measurement requires an extrapolation of the SFR, assuming that the SFR scales with the \emph{u}-band magnitude.  However, the lower the SFR in a galaxy, the higher the contribution to the flux in near-UV bands from older stars.  Thus the approximation we make would tend to overpredict the SFR in galaxies with less star-formation, leading to the type of narrow distribution of sSFRs that appear in Figure~\ref{fig:ssfrp}.

The second question is whether the slope of the correlation between \ssfrs\ and HRs is compatible with previous works.  Figure~\ref{fig:ssfrs} could be seen as being in contrast to the results of \citet{Lampeitl10} which, using photometric host-galaxy properties of SNe Ia from the SDSS-SNS, found a difference in HRs between SNe Ia in passive and SFG of $\sim0.1$ magnitude.  However, what we are measuring is the correlation with HR \emph{within} the group of all SFG; a more accurate comparison with \citet{Lampeitl10} would be the mean HR of all SNe Ia in SFGs (plotted in Figure~\ref{fig:ssfrs}) compared with that of SNe Ia found in passive galaxies.  We define passive galaxies as those hosts where the H$\alpha$ line is observed with S/N $<$ 10, and which fail at least one of the emission-line cuts described in Section~\ref{sec:late}.  For the 15 passive host galaxies with a $2\sigma$ upper limit on sSFR below that of the lowest sSFR in Figure~\ref{fig:ssfrs}, we find an average HR of $-0.037\pm0.024$.  Compared to the average HR of $0.031\pm0.012$ in our SFG sample, this is a difference of $\approx2.5\sigma$ and in agreement with \citet{Lampeitl10}.  Additionally, if we split the SFGs into equally populated upper and lower bins, we find an average HR of $0.111\pm0.018$ in the higher sSFR bin and $-0.030\pm0.017$ in the lower sSFR one, the latter of which is consistent with what we find in passive galaxies.  The similar HRs we find in both passive hosts and our lower bin of sSFR demonstrates that the linear fit plotted in Figure~\ref{fig:ssfrs}, while a good representation of emission-line galaxies, does not capture the flattening of the trend at low sSFRs.

\begin{deluxetable}{l cccccc}
\tablewidth{0pt}
\tabletypesize{\scriptsize}
\tablecaption{Significance of HR correlations with SFR and sSFR\label{tb:sfr}}
\tablehead{
  \colhead{Sample} &
  \multicolumn{2}{c}{\ssfrs\ } &
  \multicolumn{2}{c}{\ssfrp\ } &
  \multicolumn{2}{c}{\sfrp\ } \\
  & \salt & \mlcs & \salt & \mlcs & \salt & \mlcs
}
\startdata 
SFG             &  $0.1\%(3.1\sigma)$  &  $<0.1\%(3.7\sigma)$  &  12.4\%(1.1$\sigma$)  &  5.1\%(1.6$\sigma$)  &  19.9\%(0.8$\sigma$)  &  12.9\%(1.1$\sigma$)   \\
SFG+Composites  &  $0.1\%(3.2\sigma)$  &  $<0.1\%(4.4\sigma)$  &  11.4\%(1.2$\sigma$)  &  0.6\%(2.6$\sigma$)  &  21.2\%(0.8$\sigma$)  &  47.4\%($<$0.1$\sigma$)  \\
\enddata
\tablecomments{Entries are the percent of slopes in the posterior distribution of LINMIX with sign opposite that of the best-fit value, with the significance of the deviation from 0 of the best fit in parentheses.  The sample sizes differ between \sfrp, \ssfrp, and \ssfrs.  For example, `SFG' for \ssfrp\ requires a photometric host mass and is limited to fiber-based spectra only, unlike \ssfrs.}  
\end{deluxetable}
\subsection{Completeness}
\label{sec:res-complete}
We have obtained spectra for nearly all of the host galaxies of SNe Ia at redshifts $z<0.15$, covering the full range of host-galaxy absolute magnitudes.  Combined with the completeness of the SDSS-SNS at these redshifts, we expect our study to be nearly free of selection bias with respect to both SN and host-galaxy parameters.  However, having observed spectra for all of host galaxies does not mean that we are able to derive useful quantities from them all.  A given galaxy spectrum can fail our emission-line S/N cuts for two reasons:  either it is an elliptical galaxy and lacks these lines, or it is a low-luminosity galaxy.  Only the latter group is important, as a `complete' sample for our analysis by definition would exclude non-SFG.  We lose some low-luminosity host-galaxies due to our data cuts, but the Gemini spectra provide us with a high S/N sample of intrinsically faint hosts to prevent our final analysed sample from being skewed.

\begin{figure}[!htp]
  \centering
  \includegraphics[width=1.0\textwidth]{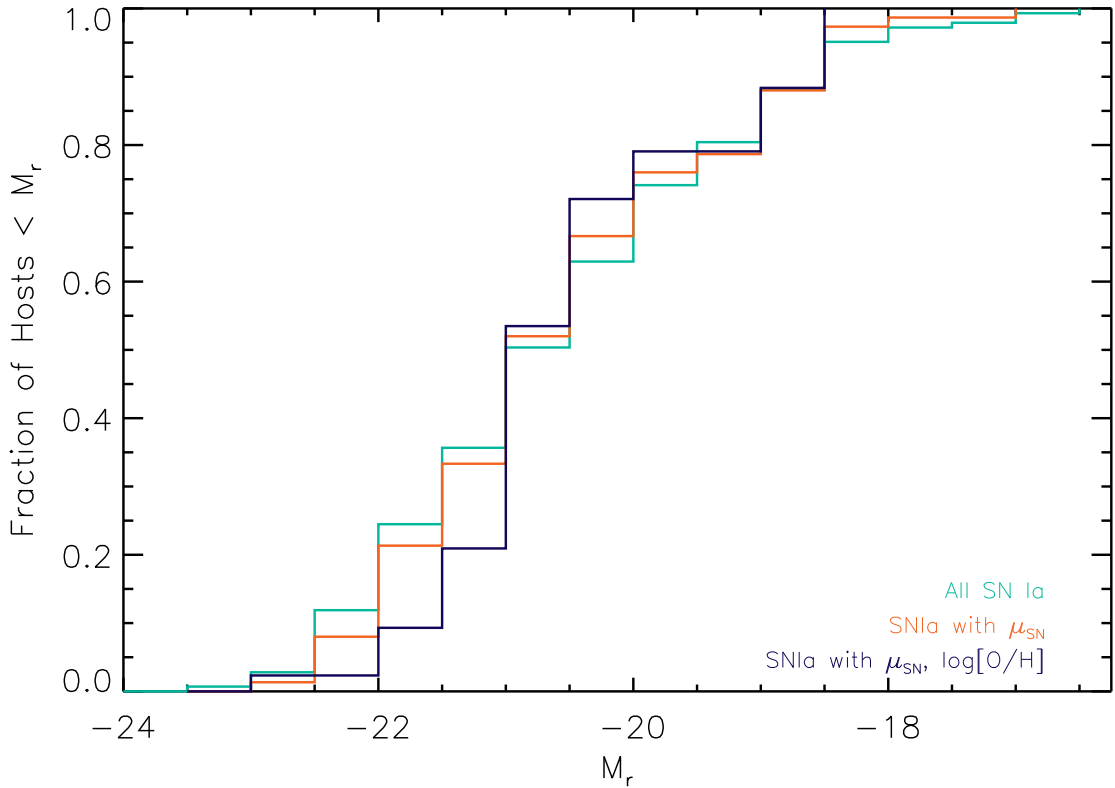}
  \caption{Cumulative distribution function in dereddened \emph{r}-band absolute magnitude for host galaxies of 1) All SNe Ia, 2) SNe Ia that have distance modulus measurements $\mu_{\textrm{SN}}$ based on \salt, and 3) SNe Ia with both a $\mu_{\textrm{SN}}$ and a measurement of the host gas-phase oxygen abundance, based on emission-line diagnostics.  All samples are for $z<0.15$.  The Kolmogorov-Smirnov test does not disfavor sample (3) - from which our final anaylsis is done - being drawn from the same distribution as sample (2).  The differences which do appear between the samples are explained in Section \ref{sec:res-complete}.
    \label{fig:ks}}
\end{figure}

Figure~\ref{fig:ks} shows the cumulative distribution function (CDF) of SNe Ia host-galaxy \emph{r}-band absolute magnitudes for all Spec Ia and Phot Ia SNe at $z<0.15$.  All galaxy magnitudes were obtained from the SDSS CAS, then de-reddened using the dust maps of \citet{Schlegel} and k-corrected using the code {\tt kcorrect v4.1.4} \citep{kcorrect}.  Included in the plot is the host-galaxy CDF for all SNe Ia in this redshift range that pass the selection requirements for having a distance modulus measurement; it is clear that this cut introduces no bias into the host magnitude distribution.  Also plotted is the sample which we draw our analysis from, the host galaxies that pass our S/N and AGN cuts (see Section~\ref{sec:late}).  We have far fewer bright galaxies, where M$_{r}<-21.5$, but this is primarily due to the fact that many of these galaxies are passive.  At the low-luminosity end, we are missing useful spectra for the faintest 5\% of host-galaxies.  This includes 4 of the 5 faintest hosts, 3 of which we observed with Gemini but were still of too low S/N for a metallicity measurement.  A Kolmogorov-Smirnov test shows a 23\% chance that the final host-galaxy magnitude distribution which we use in our analysis is drawn from the same sample as the sample which passes all SNANA cuts.  Considering that we know of and expect the differences that do exist between the two distributions, it is reasonable to state that the host-galaxy properties in this paper approximate that of an unbiased sample. 

\section{Discussion}
\label{sec:discuss}

We have examined correlations between spectroscopic host-galaxy properties in a nearly unbiased sample with SN Ia Hubble Residuals for the first time (see \citealp{Konishi} for a recent analysis of a larger but biased sample with SDSS SNe; \citealp{Gupta} for an unbiased sample with multi-wavelength host-galaxy properties, also using SDSS SNe).  We find evidence at $>4\sigma$ of higher metallicity galaxies hosting overluminous SNe Ia (after light-curve corrections are applied) when compared to lower metallicity hosts, with a difference in the mean HR of $\sim0.1$ magnitudes.  However, the number of SNe Ia for which we have host-galaxy observations is insufficient to strongly constrain the nature of the relationship between the HR and metallicity, as evidenced by the much lower significance of a correlation returned by the LINMIX package.  We also find at $>3\sigma$ confidence evidence of galaxies with low sSFRs hosting overluminous SNe, with a difference of $\sim0.1$ magnitudes between passive and star-forming hosts.  

The observed HR correlations are in the same direction as expected on theoretical grounds \citep[e.g,][though larger in magnitude]{KRW}, so it is possible to take this as evidence of the overluminosity (after light-curve corrections) of SNe Ia from metal-rich progenitors.  It is important to keep in mind, however, the difficulty in associating host-galaxy measurements with the properties of a SNe progenitors.  The metallicity measurement used in this paper is indicative of the gas-phase oxygen abundance, which is most closely associated with episodes of recent star formation.  While this should be strongly correlated with core-collapse SN progenitors, it is less clear how well this measurement relates to the older, lower mass progentiors of SNe Ia.  In addition, whether we use the `global' or `full' sample in our analysis, the metallicity we are measuring is averaged over a large physical area, and in many cases is not derived from any galaxy light from the region where the SN occured.  A useful future study would be centered not on the galaxy but the SN location itself; though given the long duration between the formation of the progenitor star and the detonation of the evolved white dwarf, it is uncertain that the location of the explosion is the region where the progenitor formed.

Supposing that the SFR we measure is representative of the environment of the SN progenitor, what does a correlation between this quantity and SNe Ia HRs imply?  Several studies have shown evidence of there being two populations of SNe Ia: `prompt' and `delayed', the monikers referring to the time elapsed since the formation of the progenitor \citep{Mannucci05,Mannucci06,SB05,Sullivan06}.  It is possible, then, that the galaxies we observe to have high sSFR are, on average, hosts of the `prompt' population, while galaxies with a low sSFR are hosts to the `delayed' population.  However, the diagnostic which we use to measure SFR tells us about only the past $\sim 10^7$~years, which is below the timescale on which even the `prompt' SNe Ia progenitors transition from formation to explosion.  

It is also possible that the sSFR is simply acting as a tracer of metallicity, meaning that the correlation which we measure between HRs and sSFR is really one of HRs and progenitor metallicity.  It is well known that for a given galaxy mass, gas-phase metallicity decreases as SFR increases \citep[see, eg.][]{Mannucci10}.  Thus, a positive HR--sSFR correlation is analogous to a negative HR--Z relationship, which are shown in Sections~\ref{sec:res-metal} and \ref{sec:res-ssfr}.  Given the small number of host galaxies we analyze in this work, we are unable to determine whether our results are a result of this degeneracy, or whether there are two distinct trends of age and metallicity with HR.  

We note that the importance of the question, `Is the measured galaxy metallicity strongly related to the progenitor metallicity,' has led to it recently being addressed theoretically by \citet{BB11}.  They find that the host metallicity \emph{does} provide a good estimate of the SN metallicity in the case of an actively star-forming galaxy, which are the type of hosts that have been analyzed in this work.

Given that we are using integrated galaxy spectra in an attempt to constrain properties of a single star - one which has since ceased to exist - it is inevitable that there be caveats and uncertainties regarding the precise relationship between the observable and the object of interest.  But the effort is worthwhile, as the origin of the intrinsic error in the Hubble Diagram and any bias it may induce as a function of redshifts will contribute a significant systematic uncertainty that upcoming wide, deep surveys (DES, LSST) will face in the pursuit of constraining cosmology with SNe Ia.  

\acknowledgements
C.D. and R.G. wish to thank Jennifer Mosher and John Fischer for helpful discussion regarding analysis and spectral reduction techniques.  

Funding for the SDSS and SDSS-II has been provided by the Alfred P. Sloan Foundation, the Participating Institutions, the National Science Foundation, the U.S. Department of Energy, the National Aeronautics and Space Administration, the Japanese Monbukagakusho, the Max Planck Society, and the Higher Education Funding Council for England. The SDSS Web Site is http://www.sdss.org/.  The SDSS is managed by the Astrophysical Research Consortium for the Participating Institutions. The Participating Institutions are the American Museum of Natural History, Astrophysical Institute Potsdam, University of Basel, University of Cambridge, Case Western Reserve University, University of Chicago, Drexel University, Fermilab, the Institute for Advanced Study, the Japan Participation Group, Johns Hopkins University, the Joint Institute for Nuclear Astrophysics, the Kavli Institute for Particle Astrophysics and Cosmology, the Korean Scientist Group, the Chinese Academy of Sciences (LAMOST), Los Alamos National Laboratory, the Max-Planck-Institute for Astronomy (MPIA), the Max-Planck-Institute for Astrophysics (MPA), New Mexico State University, Ohio State University, University of Pittsburgh, University of Portsmouth, Princeton University, the United States Naval Observatory, and the University of Washington.

Funding for SDSS-III has been provided by the Alfred P. Sloan Foundation, the Participating Institutions, the National Science Foundation, and the U.S. Department of Energy. The SDSS-III web site is http://www.sdss3.org/.  SDSS-III is managed by the Astrophysical Research Consortium for the Participating Institutions of the SDSS-III Collaboration including the University of Arizona, the Brazilian Participation Group, Brookhaven National Laboratory, University of Cambridge, University of Florida, the French Participation Group, the German Participation Group, the Instituto de Astrofisica de Canarias, the Michigan State/Notre Dame/JINA Participation Group, Johns Hopkins University, Lawrence Berkeley National Laboratory, Max Planck Institute for Astrophysics, New Mexico State University, New York University, Ohio State University, Pennsylvania State University, University of Portsmouth, Princeton University, the Spanish Participation Group, University of Tokyo, University of Utah, Vanderbilt University, University of Virginia, University of Washington, and Yale University.

The Apache Point Observatory 3.5-meter telescope is owned and operated by the Astrophysical Research Consortium.  We thank the observatory director, Suzanne Hawley, and site manager, Bruce Gillespie, for their support of this project.  Observations at the ESO New Technology Telescope at La Silla Observatory were made under programme {\small ID}s 77.A-0437, 78.A-0325, and 79.A-0715. Based on observations obtained at the Gemini Observatory, which is operated by the Association of Universities for Research in Astronomy, Inc., under a cooperative agreement with the NSF on behalf of the Gemini partnership: the National Science Foundation (United States), the Science and Technology Facilities Council (United Kingdom), the National Research Council (Canada), CONICYT (Chile), the Australian Research Council (Australia), Minist\'{e}rio da Ci\^{e}ncia e Tecnologia (Brazil) and Ministerio de Ciencia, Tecnolog\'{i}a e Innovaci\'{o}n Productiva (Argentina).

\clearpage
\end{document}